\documentclass[aps,prb,twocolumn]{revtex4}
\usepackage{graphicx,epsf}
\usepackage{amsmath}

\newcommand{\ybco}{YBa$_2$Cu$_3$O$_{6+\delta}$}

\newcommand{\lsco}{La$_{2-x}$Sr$_x$CuO$_4$}
\newcommand{\bscco}{Bi$_2$Sr$_2$CaCu$_2$O$_{8+\delta}$}
\newcommand{\lbco}{La$_{2-x}$Ba$_x$CuO$_4$}

\newcommand{\ccoc}{Ca$_{2-x}$Na$_x$CuO$_2$Cl$_2$}

\begin{document}

\title{
Electronic properties of disordered valence-bond stripes in cuprate superconductors
}

\author{Matthias Vojta}
\affiliation{
Institut f\"ur Theoretische Physik, Universit\"at zu K\"oln,
Z\"ulpicher Stra\ss e 77, 50937 K\"oln, Germany
}
\date{Sep 17, 2008}

\begin{abstract}
We calculate single-particle properties of short-range ordered stripe states,
using Monte-Carlo simulations of collective charge-density wave (CDW)
order parameters coupled to fermions on a 2d square lattice.
For superconducting bond-centered stripes with a $d$-wave form factor,
we find a valence-bond ``glass''
which coexists with low-energy quasiparticles featuring interference phenomena,
in agreement with recent scanning-tunneling-microscopy (STM) measurements
on underdoped \bscco\ and \ccoc.
Together with earlier work,
our calculations provide a link between CDW signatures seen in STM and
those in magnetic neutron scattering.
\end{abstract}

\maketitle


\section{Introduction}

Charge-density wave (CDW) phenomena have been detected in a number
of superconducting cuprates.
Most prominent are the uni-directional spin and charge modulations,
termed ``stripes'',
in \lbco\ and \lsco\ (with Nd or Eu co-doping),
being strongest near 1/8th doping.\cite{jt95,jt97,yamada98,pnas,abbamonte}
In other cuprates, notably \bscco\ and \ccoc, scanning tunneling microscopy (STM)
measurements have found signatures of short-range charge
order.\cite{hana,ali,mcelroy,kohsaka}.

Remarkably, the existence of stripe states was postulated in early theory
work on the Hubbard model,\cite{zaanen,schulz,machida}
far before experimental indications for such phases were found.
Later on, ideas of frustrated phase separation as driving force of
stripe formation were worked out in detail,\cite{ek94}
and CDW quantum criticality was proposed as source of both non-Fermi liquid behavior
and superconductivity.\cite{castellani}
For the vast number of subsequent theoretical activities we refer to
the review articles Ref.~\onlinecite{stevek,ahcn04}.

While the role of charge order for the overall properties of cuprates
is under debate, a plausible hypothesis is that tendencies toward
charge ordering are common to underdoped cuprates.\cite{ek94,castellani,jan,stevek,ahcn04,doug}
Even compounds not displaying long-range order are influenced
by the proximity to a charge-ordered state.
In particular, impurities will act as random-field pinning centers
for the collective charge modes, leading to static short-range order
(as observed in STM).\cite{stevek,maestro,robertson}
Moreover, charge order will influence the magnetic excitations,
believed to be the pairing glue:\cite{natphys}
It was recently shown \cite{vvk} that short-range-ordered stripes
give rise to an ``hour-glass'' magnetic spectrum,
very similar to that observed in neutron scattering experiments both
on \lbco\cite{jt04} and \ybco.\cite{hayden}

While neutron and X-ray scattering were used to detect superstructure
modulations from long-range charge order,\cite{jt95,jt97,abbamonte}
there is relatively little information on the electronic structure
of stripe states.
Both STM and photoemission indicate the presence of
coherent, gapless nodal quasiparticles (QP) in (1,1) direction,
whereas antinodal QP in (1,0) direction are rather incoherent
and likely dominated by charge ordering.\cite{mcelroy,shen,hana07}
For the compound \lbco, a $d$-wave-like gap was recently reported,\cite{valla}
which may be attributed to static stripes or to fluctuating superconductivity.\cite{li}

In this paper, we present a detailed study of local electronic properties
of disordered stripe states in cuprates,
using a CDW order-parameter approach
plus a mean-field theory for the single-particle dynamics.
A central ingredient is the $d$-wave-like form factor of the
charge order,\cite{MVOR}
which causes the modulations to be located primarily on Cu-O-Cu bonds
instead of on Cu sites.
Our results reproduce central features of the STM data of Refs.~\onlinecite{kohsaka,hana07}.
As we employ the {\em same} model for the collective CDW modes
as in Ref.~\onlinecite{vvk}, used there to calculate spin
excitations in the presence of disordered stripes,
our results provide a link between different probes of stripe physics.

The remainder of the paper is organized as follows:
In Sec.~\ref{sec:model} we describe the employed model together with
the approximations and their physical background.
Sec.~\ref{sec:res} presents the main numerical results,
with focus on describing the STM data of Refs.~\onlinecite{kohsaka,hana07}.
A discussion and conclusion closes the paper.


\section{Phenomenological modelling}
\label{sec:model}

Our phenomenological model consists of coupled CDW fluctuations and electrons, with the
action ${\cal S} = {\cal S}_\psi + {\cal S}_c + {\cal S}_{c\psi}$.
To account for the strong commensuration effects observed experimentally,
all fields will be defined for discrete lattice coordinates.\cite{zachar98}

\subsection{Lattice CDW order-parameter theory}

The CDW part ${\cal S}_\psi$ captures the tendency toward stripe ordering
and is identical to that of Ref.~\onlinecite{vvk}:
Two complex fields $\psi_{x,y} ({\vec r}, \tau)$
represent the amplitude of horizontal and vertical stripe order at wavevectors
${\vec K}_{x,y}$, such that the real field
$
Q_x ({\vec r} ) = {\rm Re}\,\psi_x ({\vec r}) e^{i {\vec K}_x \cdot {\vec r}}
$
(similarly for $Q_y$) measures the modulation of both the
charge density and bond order (i.e., kinetic energy or pairing amplitude),
for $\vec r$ on sites and bonds, respectively.
Then, $\delta\rho({\vec r}_j) = Q_x + Q_y$ is the deviation of the local hole density
from its spatial average.
We restrict our attention to
${\vec K}_{x}\!=\!(\pi/2,0)$ and ${\vec K}_y\!=\!(0,\pi/2)$, i.e.,
a charge modulation period of 4 lattice spacings.\cite{ali,hana}
The complex phase of $\psi_{x,y}$ represents the sliding degree of freedom
of the density wave.

Fluctuations of the charge order are described by a $\psi^4$-type theory ${\cal S}_\psi$
for the O(4) field $\psi = (\psi_x,\psi_y)$.
The precise form of ${\cal S}_\psi$ will determine the character of the
fluctuations (amplitude vs. phase).
The STM data of Ref.~\onlinecite{hana}, with modulations present everywhere in real space,
point toward small amplitude fluctuations;
in addition, the calculated spin-fluctuation spectra of Ref.~\onlinecite{vvk} were only
compatible with experiment under the assumption of dominant phase fluctuations.
Hence, we employ
\begin{eqnarray}
&&\mathcal{S}_{\psi} = \int \!d\tau \! \sum_i \Bigl[
\left| \partial_\tau \psi_{ix} \right|^2 +
\left| \partial_\tau \psi_{iy} \right|^2 +
s_x |\psi_{ix}|^2 + s_y |\psi_{iy}|^2 \nonumber\\
&&+
 c_{1x}^2 | \psi_{ix}-\psi_{i+x,x} |^2 +
 c_{2x}^2 | \psi_{ix}-\psi_{i+y,x} |^2 \nonumber\\
&&+
 c_{1y}^2 | \psi_{iy}-\psi_{i+x,y} |^2 +
 c_{2y}^2 | \psi_{iy}-\psi_{i+x,y} |^2
\nonumber \\
&&+
u_1 \psi_i^4 + u_2 \psi_i^6
+ v |\psi_{ix}|^2 |\psi_{iy}|^2 \nonumber\\
&&+ w \left( \psi_{ix}^4 \!+\! \psi_{ix}^{\ast 4}
\!+\! \psi_{iy}^4 \!+\! \psi_{iy}^{\ast 4} \right) \Bigr]
\label{spsi}
\end{eqnarray}
with $\psi_{ix}\equiv \psi_x(\vec{r}_i)$ and $\psi_i^2\!\equiv\!|\psi_{ix}|^2\!+\!|\psi_{iy}|^2$.
A combination of $u_1\!<\!0$ and $u_2\!>\!0$ suppresses amplitude fluctuations
of $\psi$.
The quartic $v |\psi_x|^2 |\psi_y|^2$ term regulates the repulsion or
attraction between horizontal and vertical stripes;
we shall mainly employ $v\!>\!0$ leading to stripe-like order
(whereas $v\!<\!0$ results in checkerboard structures).
The phase-sensitive $w$ term provides commensurate pinning and
selects bond-centered (instead of site-centered)
stripes\cite{kohsaka,vvk} for $w\!>\!0$.


\subsection{Fermions}

To calculate electronic properties in the presence of collective charge modes,
we start from a BCS model of fermions on the square lattice of Cu atoms:\cite{notation}
\begin{eqnarray}
{\cal S}_c &=& \int \! d\tau \! \sum_{{\vec k}} \left[
\bar{c}_{{\vec k}\sigma} (\partial_\tau\!+\!\epsilon_{\vec k}\!-\!\mu) c_{{\vec k}\sigma} +
\Delta_{\vec k} (c_{{\vec k}\uparrow} c_{{-\vec k}\downarrow} + c.c.)
\right]
\nonumber\\
\label{sc}
\end{eqnarray}
where summation over spin indices $\sigma$ is implied.
The single-particle dispersion consists of hopping to
first, second, and third neighbors, with
$t\!=\!-0.15$ eV, $t'\!=\!-t/4$, $t''\!=\!t/12$.
The chemical potential is $\mu\!=\!-0.12$ eV, leading to
a hole doping of $\approx 11\%$.
The pairing is of $d$-wave type, $\Delta_{\vec k} = \Delta_0 (\cos k_x\!-\!\cos k_y)$
with $\Delta_0\!=\!24$ meV.

The coupling to the collective CDW fields $Q_{x,y}$ reads
\begin{eqnarray}
{\cal S}_{c\psi} &=& \int d\tau \sum_i \Big[
\kappa_1 Q_x({\vec r}_i) \bar{c}_{i\sigma} c_{i\sigma} \nonumber\\
+&&\!\!\!\!\!\!\!\!\!\!
\big(
\kappa_2 Q_x({\vec r}_{i+x/2}) \bar{c}_{i\sigma} c_{i+x,\sigma} \,+\,
\kappa_3 Q_x({\vec r}_{i+y/2}) \bar{c}_{i\sigma} c_{i+y,\sigma} \nonumber\\
+&&\!\!\!\!\!\!\!\!\!\!
\kappa_4 Q_x({\vec r}_{i+x/2}) c_{i\uparrow} c_{i+x\downarrow} +
\kappa_5 Q_x({\vec r}_{i+y/2}) c_{i\uparrow} c_{i+y\downarrow} \!+\! c.c.
\big) \nonumber\\
+&&\!\!\!\!\!\!\!\!\!\!
[x \leftrightarrow y] \Big]
\label{scpsi}
\end{eqnarray}
with $Q_x({\vec r}_{i+x/2})=[Q_x({\vec r}_{i})+Q_x({\vec r}_{i+x})]/2$.
The coupling constants $\kappa_{1\ldots5}$ decide about the electronic
struture of the CDW state, by implementing modulations of
charge densities and bond kinetic and pairing energies.
In the simplest picture, stripes correspond to modulations in the
on-site charge densities.
Those are induced by $\kappa_1$ and
lead to a nearly $\vec k$-independent ($s$-wave) CDW form factor
$\phi_2({\vec k}) = \langle c_{{\vec k+\vec K},\sigma}^\dagger  c_{{\vec k}\sigma}
\rangle$.\cite{MVOR,notation}
However, local ordering can instead be dominated by physics on Cu-O-Cu {\em bonds}:
Stripe formation is driven by the competition between kinetic and magnetic
energies, both living on bonds.\cite{ssrmp,sr,vs}
We have recently argued\cite{MVOR} that such a bond-dominated stripe state
will have modulations in $\langle c_{i\sigma}^\dagger c_{i+\Delta,\sigma}\rangle$
with locally different signs on horizontal and vertical bonds,
implying a strong $d$-wave component of $\phi_2({\bf k})$,
see Fig. 1 of Ref.~\onlinecite{MVOR}.
Modulations on bonds are induced by $\kappa_{2\ldots5}$,
with the $d$-wave character encoded, e.g., in $\kappa_2=-\kappa_3$.

A few remarks are in order:
In the advocated model, Eqs.~(\ref{spsi},\ref{sc},\ref{scpsi}),
correlation effects are included via
$\epsilon_{\vec k}$ being a renormalized quasiparticle dispersion and
via $Q_{x,y}$ representing collective CDW tendencies,
while genuine Mott physics is absent.
Dispersion renormalizations are standard in mean-field theories
of correlated electrons; here we refrain from a self-consistent
calculation of the dispersion and instead use plausible hopping
parameters extracted from photoemission experiments.
The separation of degrees of freedom into quasiparticles and
collective CDW fields, both with full spatial or momentum dependence,
is phenomenological and cannot be rigorously justified.
However, e.g., in the context of electrons interacting with antiferromagnetic fluctuations,
this has proven to be a fruitful route of investigation.\cite{chub}


\subsection{Observables}

STM experiments determine the spatially resolved local density of states
(LDOS), $\rho(\vec{r},E)$, up to an $\vec{r}$-dependent tunnel matrix element
(which depends on the set-point conditions.\cite{kohsaka,hana07})
To separate physical modulations from set-point effects, the LDOS ratios
\begin{eqnarray}
Z(\vec{r},E) &=& \frac{\rho(\vec{r},E)}{\rho(\vec{r},-E)}, \nonumber\\
R(\vec{r},E) &=& \frac{\int_0^E d\omega\rho(\vec{r},\omega)}{\int_{-E}^0 d\omega \rho(\vec{r},\omega)}
\end{eqnarray}
have been used.
In a weakly doped Mott insulator, both $Z$ and $R$
(measuring spectral particle--hole asymmetry)
can be shown to be proportional to the hole density.\cite{pwa_as,mohit_as}
In Ref.~\onlinecite{kohsaka}, spatial modulations were
observed in $R(\vec{r},E)$.
In the following, we shall assume that these reflect
modulations in the hole density.\cite{mohit_as,ZR_foot}

\subsection{Perfect CDW order}

Perfectly ordered CDW states are described by ${\cal S}_c+{\cal S}_{c\psi}$,
with $\psi_{x,y}$ taken to be constant.
From the diagonalized fermionic Hamiltonian all electronic properties
can be obtained.

\begin{figure}[!t]
\epsfxsize=3.4in
\centerline{\epsffile{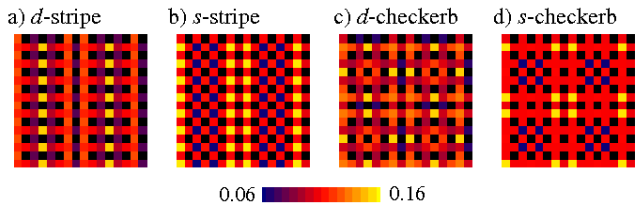}}
\caption{
Local densities for ordered CDW states,
calculated from ${\cal S}_c+{\cal S}_{c\psi}$ for constant $\psi$.
The figures show {\em both} site and bond densities,
$\langle 1-c_{i\sigma}^\dagger c_{i\sigma}\rangle$ and
$\langle c_{i\sigma}^\dagger c_{i+\Delta,\sigma} + h.c.\rangle$
\cite{bondfoot} for $8^2$ Cu sites;
black squares are centers of CuO$_2$ plaquettes.
a) Stripes with $\kappa_1\!=\!0$, $\kappa_2\!=\!-\kappa_3\!=\!0.03$, $-\kappa_4\!=\!\kappa_5\!=\!0.007$ ($d$-wave).
b) Stripes with $\kappa_1\!=\!0.022$, $\kappa_{2\ldots 5}\!=\!0$ ($s$-wave).
c) Checkerboard [$d$-wave as in a)].
d) Checkerboard [$s$-wave as in b)].
}
\label{fig:ord}
\end{figure}

Sample results for the real-space densities of different types of CDW
are displayed in Fig.~\ref{fig:ord}.
Here, $\psi_x=(1+i)/\sqrt{2}$, $\psi_y=0$ for the bond-centered stripes in panels a) and b),
while $\psi_x=\psi_y=(1+i)/\sqrt{2}$ for the checkerboards in panels c) and d).
The couplings $\kappa$ were taken to induce $s$-wave-like [panels b) and d)]
or $d$-wave-like [panels a) and c)] modulations, and the overall $\kappa$
amplitude was chosen such that the resulting modulation of fermionic densities
is about 30-40\%.
To facilitate comparison with STM data,\cite{kohsaka} which show a strong
modulation on the {\em bonds} of the CuO$_2$ plane,\cite{beyond}
we included the bond charge densities $\langle c_{i\sigma}^\dagger c_{i+\Delta,\sigma} + h.c.\rangle$
(i.e. kinetic energies) in Fig.~\ref{fig:ord} --
those are shown in between the square-lattice sites.\cite{bondfoot}
Clearly, Fig.~\ref{fig:ord}a with $d$-wave stripes is most compatible
with experiment.\cite{kohsaka}

\subsection{Pinning and adiabatic approximation}

The treatment of $\mathcal{S}_c\!+\!\mathcal{S}_{\psi}\!+\!\mathcal{S}_{c\psi}$
requires additional input.
Pinning is important especially in the disordered phase of $\mathcal{S}_{\psi}$:
Quenched disorder (e.g. from dopant impurities) acts as a random field
and renders {\em static} a short-range ordered stripe configuration.
In such a situation, the electronic properties can be approximately
calculated by diagonalizing ${\cal S}_c+{\cal S}_{c\psi}$ for fixed
static configurations of $\psi_{x,y}$.
We generate these from classical lattice Monte Carlo (MC)
simulations of $\mathcal{S}_{\psi}$, using a standard Metropolis algorithm
at a finite effective temperature ($T=1$) in a regime where
the stripe correlation length is of order $\xi \approx 10$.\cite{pinfoot}

The numerical procedure parallels that of the adiabatic approximation of
Ref.~\onlinecite{vvk}, with the difference that the ingredient of pinning
is crucial to obtain a static signal in STM.
(The presence or absence of pinning was irrelevant to the finite-frequency
spin fluctuations described in Ref.~\onlinecite{vvk}.)
In contrast to earlier work dealing with fermionic properties in the presence
of disordered stripes,\cite{salkola}
our modelling implements the $d$-wave bond character,
and it properly describes short-range order via $S_\psi$ \eqref{spsi},
i.e., stripe segments
coexist with checkerboard domain walls.\cite{vvk,maestro,robertson}

\subsection{Choice of parameters and validity of approximation}

The parameters of $\mathcal{S}_c\!+\!\mathcal{S}_{c\psi}$, Eqs.~(\ref{sc},\ref{scpsi}),
used in our simulations are taken as in the static-stripe calculation above,
i.e., for the fermionic sector we use values for $t$, $t'$, $t''$, and $\Delta$, which
are standard in the BCS mean-field description of cuprates,
and the couplings $\kappa$ are taken as in Fig.~\ref{fig:ord}a.

The CDW part of the action, $\mathcal{S}_{\psi}$ \eqref{spsi}, is designed to capture the
complicated non-universal physics of the strongly correlated CDW formation on the lattice
scale. The combination of $s_{x,y}$, $u_1$ and $u_2$ decides about the importance of
amplitude vs. phase fluctuations of the CDW,\cite{vvk}
we have used $s_x\!=\!s_y\!=\!-4\ldots-3$, $u_1\!=\!-1.15$, $u_2\!=\!0.1$.
Choosing $v\!=\!0.2$ prefers stripes over checkerboards, but
allows for some checkerboard structure between stripe domains.\cite{vvk}
Finally, $w\!=\!0.05$ is taken for a moderate commensurate lattice pinning
toward bond-centered stripes.
The precise values of the mass $s$ and the gradient $c$ were used to tune
the CDW correlation length $\xi$, which was between 10 and 30 in our simulations.
Note that an overall scale factor in $\mathcal{S}_{\psi}$ is free
and determines the typical amplitude of $\psi$ which we have normalized to unity.
The $\mathcal{S}_{\psi}$ parameters here are identical to those used
in Figs. 1a, 2a of Ref.~\onlinecite{vvk} for the description of the spin excitations
of fluctuating stripes.
Moreover, the charge configurations generated from the MC simulations
visually match the STM results in the sense that short and medium stripe segments
coexist with checkerboard-like domain walls.
This property is robust with respect to parameter changes of 20\% and more,
provided that the correlation length $\xi$ is kept fixed.
As we employ classical MC simulations for $\mathcal{S}_{\psi}$,
with time gradients absent, the parameters cannot be translated directly
into physical energies or velocities.

Our approach assumes that a mean-field picture of both superconductivity and
charge order is a reasonable starting point for the description of cuprates.
The adiabatic approximation neglects inelastic processes and stripe dynamics,
which can be justified if the latter is slow (compared to the observed fermions),
as happens in the proximity to a CDW ordering transition.
Thus, the approximation is invalid for energies below a typical stripe fluctuation
frequency; for strong impurity pinning this scale is small or zero.\cite{pinfoot}
The quasiparticle picture in cuprates may break down at elevated energies;
as far as this happens due to inelastic physics, it is not captured by our approach
(while some elastic disorder physics is captured).

Further, we assume that dimerization and bond order are the driving forces
behind stripe ordering,\cite{vs,vvk} whereas magnetic long-range order is less
important. For simplicity, we therefore neglect both order and fluctuations
in the triplet channel.
Note that this does not mean that we ignore local-moment physics entirely,
but instead we assume that those moments form singlet valence bonds,
which is accounted for by modulated hoppings ($\kappa_{2,3}$) in $\mathcal{S}_{c\psi}$ \eqref{scpsi}.
We note that the coupling to magnetic fluctuations will contribute
to the broadening primarily of antinodal quasiparticles,\cite{soc}
but a calculation including spatial disorder and inelastic processes
is beyond the scope of the present paper.


\section{Numerical results}
\label{sec:res}

We now turn to the numerical results obtained from
$\mathcal{S}_c\!+\!\mathcal{S}_{\psi}\!+\!\mathcal{S}_{c\psi}$
for short-range ordered stripes.

\begin{figure}[!t]
\epsfxsize=3.1in
\centerline{\epsffile{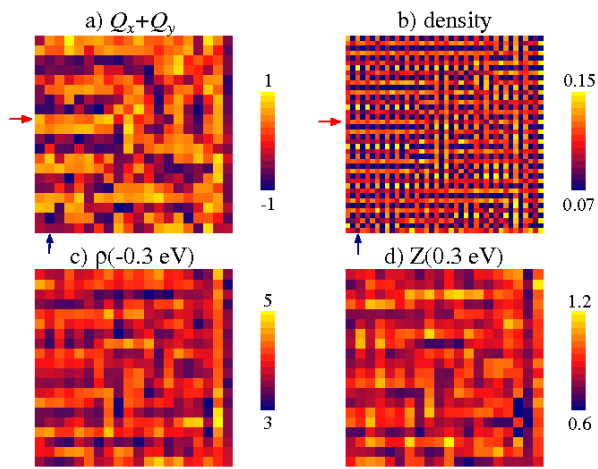}}
\caption{
Local observables for pinned short-range ordered stripes
($\kappa$ parameters as in Fig.~\ref{fig:ord}a,
correlation length $\xi\approx10$),
showing $20^2$ Cu sites of the $64^2$ sample.
a) Order parameter field $Q_x+Q_y$.
b) Local density (including bonds) as in Fig.~\ref{fig:ord}.
c) LDOS at $-0.3$ eV.
d) $Z(\vec{r},E)$ at $0.3$ eV.
The arrows mark the rows/columns along which the LDOS is shown in
Fig.~\ref{fig:spectra}.
}
\label{fig:dis1}
\end{figure}

\subsection{Local densities}

\begin{figure}[!t]
\epsfxsize=3.1in
\centerline{\epsffile{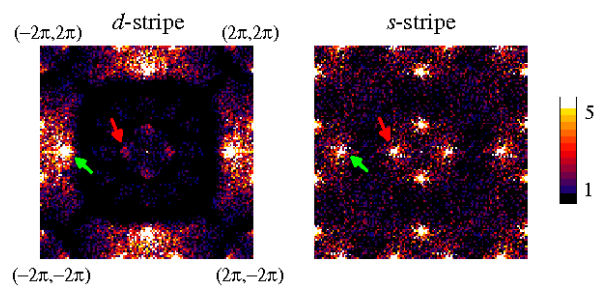}}
\caption{
Fourier-transformed local density (from sites and bonds\cite{bondfoot})
for short-range ordered stripes,
obtained from averaging 40 MC configurations on a $64^2$ lattice.
Left: $\kappa_1\!=\!0$, $\kappa_2\!=\!-\kappa_3\!=\!0.03$, $-\kappa_4\!=\!\kappa_5\!=\!0.007$ ($d$-wave).
Right: $\kappa_1\!=\!0.022$, $\kappa_{2\ldots 5}\!=\!0$ ($s$-wave).
The dark/light (red/green) arrows indicate the peaks at
$(\pi/2,0)$ and $(3\pi/2,0)$.
(The strong peaks at the zone boundary arise from artificially setting the
density to zero in the plaquette centers.)
}
\label{fig:ftdens}
\end{figure}

Fig.~\ref{fig:dis1} displays the order parameter field
$(Q_x\!+\!Q_y)$ together with the resulting fermionic
charge density, the LDOS $\rho(\vec{r},E)$, and $Z(\vec{r},E)$ at
a high energy of $0.3$ eV, for one fixed $\psi_{x,y}$ configuration\cite{pinfoot}
for a $d$-wave coupling in ${\cal S}_{c\psi}$.
The stripe modulation, being prominent on the bonds, leads
to a large contrast in Fig.~\ref{fig:dis1}b, while
the contrast in both the site-LDOS $\rho$
and $Z$ (Figs.~\ref{fig:dis1}c,d) is weaker.
(Both $\rho$ and $Z$ show strong modulations around the gap energy,
Fig.~\ref{fig:spectra} below.)
The result in Fig.~\ref{fig:dis1}b has a striking similarity
to the ``glassy'' structures in Figs. 3,4 of Ref.~\onlinecite{kohsaka}.
In particular, the modulation locally breaks the C$_4$ rotation
symmetry down to C$_2$ and is primarily located on the Cu-O-Cu
bonds.
The latter fact -- which originates in the $d$-wave form factor --
can be nicely seen in the Fourier-transformed density,
Fig.~\ref{fig:ftdens}.
Stripe order is manifest in peaks at $(\pi/2,0)$, $(0,\pi/2)$
and $(3\pi/2,0)$, $(0,3\pi/2)$, with the signal at
$(3\pi/2,0)$ being much stronger compared to $(\pi/2,0)$
(whereas for $s$-wave stripes the peaks are roughly equal
in intensity).
Again, this is in agreement with STM data,
Fig. 6 of Ref.~\onlinecite{kohsaka}.

We note that the present comparison between theory
and experiment does not easily allow to deduce the
amplitude of the actual modulations:
The only observables free of set-point effects are $Z$ and $R$.
However, a reliable calculation of these has to
has to cope with Mott physics not included in our model.\cite{ZR_foot}
A rough estimate, however, relates the experimentally observed
$R$ contrast of $\approx\pm30\%$ to a bond modulation
of similar magnitude.

\subsection{Nodal quasiparticles and quasiparticle interference}

One outstanding feature of the STM results on underdoped \ccoc\
is the presence of quasiparticle interference (QPI) features in the low-energy
spectra, in a situation where the high-energy spectra are
dominated by period-4 modulations.\cite{hana07}

\begin{figure}[!t]
\epsfxsize=3.5in
\centerline{\epsffile{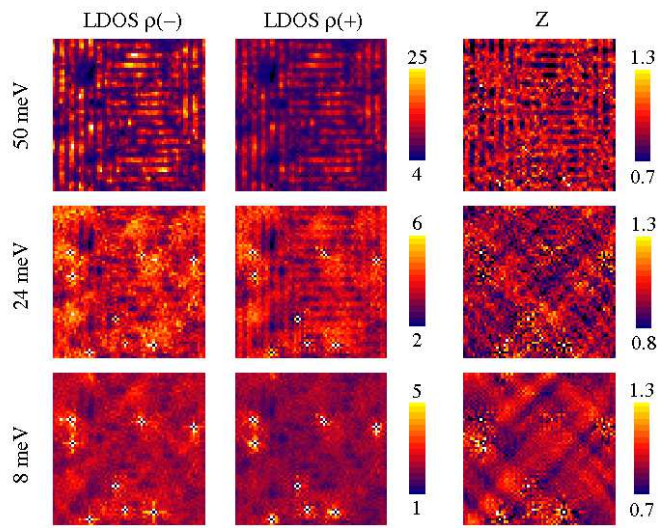}}
\caption{
LDOS $\rho(\vec{r},E)$ at negative (left) and positive (middle) bias,
together with $Z(\vec{r},E)$ (right),
for pinned short-range ordered stripes
with additional impurities on a $64^2$ lattice,
at energies 50, 24, 8 meV (from top to bottom).
$\kappa$ parameters are as in Fig.~\ref{fig:ord}a,
for details see text.
}
\label{fig:qpi}
\end{figure}

Our calculations qualitatively reproduce this physics.
For QPI to occur, we have to add realistic disorder as source of QP scattering:
Following Ref.~\onlinecite{nunner2}, we use a combination of
2\% extended potential scatterers (strength 40 meV, size 1.2),
5\% extended pairing scatterers (strength $\Delta_0$, size 1.5),
and 0.2\% pointlike unitary scatterers (strength 2.5 eV).
The real-space results of such a calculation are displayed in Fig.~\ref{fig:qpi}.
Both the LDOS and the $Z$ map at higher energies are clearly dominated
by stripe segments, whereas the signal below $\approx 25$ meV shows the typical
QPI modulations (compare e.g. Fig. 3c of Ref.~\onlinecite{hana07}).
Extracting the scattering wavevectors from the Fourier transform
of our data (not shown) is difficult due to the small system size;
the only unambigous peak is at the so-called $q_7$ wavevector
(corresponding to the diagonal modulations in $Z$ at low $E$, Fig.~\ref{fig:qpi}).

We point out two features of our results.
(i) The $Z$ map is more sensitive to QPI than the LDOS,
because, to leading order, QPI modulations at positive and negative $E$
are anti-phase, while stripe modulations are in-phase.
Nevertheless, the strong period-4 modulations seen in the experimental
low-energy LDOS \cite{hana07} are likely due to set-point effects.
(ii) Real-space localization of antinodal QP {\em cannot} be made
responsible for the loss of QPI at higher $E$.
We have calculated the inverse participation ratio (not shown)
as an indicator of localization, and have observed no
localization signatures on scales up to several $\xi$
(while these length scales are sufficient to observe QPI).

More generally, the compatibility of stripes with long-lived nodal
QP has been pointed out in the past.\cite{vs,MVOR,granath01,wavev,granath08}
For small stripe amplitude,
this already follows from the fact that the ordering wavevector
$\bf Q$ does {\em not} connect the nodal points.\cite{vs,wavev}
In our case, nodal QP survive even for {\em large} stripe amplitude
due to the $d$-wave character of the charge order
(provided that the $s$-wave component remains small).\cite{MVOR}

\begin{figure}[b]
\epsfxsize=3.2in
\centerline{\epsffile{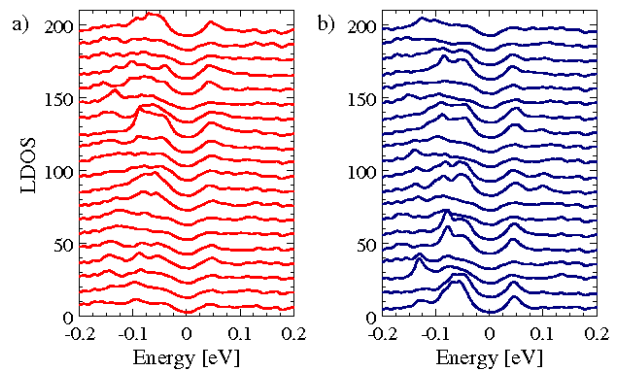}}
\caption{
LDOS spectra corresponding to the data in Fig.~\ref{fig:dis1},
along the two paths indicated by arrows in Figs.~\ref{fig:dis1}a,b.
Spectra have been broadened by 7 meV and shifted for clarity.
}
\label{fig:spectra}
\end{figure}

Within our simulations,
the survival of coherent nodal QP in the presence of disordered stripes
is also seen in the LDOS spectra in Fig.~\ref{fig:spectra}, taken along two different
line cuts indicated in Fig.~\ref{fig:dis1}.
While strong inhomogeneities occur at elevated energies, in particular
near the gap energy (note the period-4 modulation in Fig.~\ref{fig:spectra}b around $-50$ meV),
the low-energy part of the LDOS is essentially homogeneous,\cite{granath08}
again in striking similarity to STM data.\cite{mcelroy,kohsaka}
(A detailed comparison of our spectra with experiment reveals several
differences, which we believe to be related to Mott physics not captured here.)


\section{Conclusions}

We have determined electronic properties of short-range ordered
stripe states, coexisting with superconductivity.
Agreement with salient features of STM experiments,
in particular stripy LDOS modulations at elevated energies
coexisting with QP interference at low energies,
is found for valence-bond stripes with $d$-wave-like form factor,
singling out a specific mean-field plus stripe disorder model.

As the {\em same} collective-mode description was used earlier
to model magnetic excitations in the presence of fluctuating or disordered
stripes,
our calculations give a unified account of stripe signatures
seen in STM and in neutron scattering, and strongly indicate that
similar physics underlies the modulated states observed
in different underdoped cuprates.

Very recent STM experiments\cite{extinct} indicate the quasiparticle interference
disappears not only at high energies, but at a very specific location in
momentum space, approximately at the boundary of the antiferromagnetic
Brillouin zone.
Such a feature is not part of the present theory, and likely requires
to take into account either antiferromagnetic fluctuations or
other precursors of strong Mott physics.

\acknowledgments

We thank J. C. Davis, H. Takagi, and A. Yazdani for discussions,
and R. K. Kaul, S. Sachdev, T. Vojta, and A. Wollny
for collaborations on related work.
This research was supported by the DFG through
SFB 608 (K\"oln) and the Research Unit FG 538.



\begin{thebibliography}{}
\bibitem{notation}
We use the standard fermionic notation where $c^\dagger,c$ are
canonical operators, whereas $\bar{c},c$ denote the corresponding
Grassmann numbers in the coherent-state path integral.
\bibitem{ZR_foot}
In our calculations, the quantity $Z$ is sensitive to
stripe modulations and QPI, but does not measure ``Mottness'' due to the
mean-field character of $S_\psi+S_{c\psi}$.
\bibitem{beyond}
Microscopics beyond one-band models may be required
to obtain a full picture of bond order in cuprates.
\bibitem{pinfoot}
The real-space results in this paper have been calculated from a single
MC configuration of ${\cal S}_\psi$ each.
We have also performed simulations of ${\cal S}_\psi$ with different
pinning potentials included (whose microscopic form, however, is not known).
The obtained results are similar to the ones shown here
(provided that $\xi$ is comparable).
\end{thebibliography}
\end{document}